# Scaling Issues for AlGaN/GaN HEMTs: Performance Optimization via Devices Geometry Modelling

S. Russo, and A. Di Carlo

*Abstract*—**The potential barrier between source and gate in HEMTs and between source and channel in MOSFET controls the current output and the velocity injection of electrons in the channel [1], [2]. In non self aligned structures the electric field behavior along the channel, for fixed applied voltages, is determined by the contacts positions. Anyway, in GaAs based HEMTs, the geometry of the device appears to be not determinant for the output current due to saturation effects. On the other hand, the GaN based technology still offers the possibility to enhance devices output current handling contacts distances. In this paper we will present Monte Carlo simulations results which show how a downscaling of the Source to Gate distance could improve the device performances inducing an higher potential barrier between source and gate.**

*Index Terms*—**Gallium Nitride, HEMT, Electric Field, saturation current.**

## I. Introduction

THE GaN based HEMTs are one of the most promising technology for high power, high frequency applications. This is mainly due to the wide band-gap, the high peak saturation velocity of the Gallium Nitride semiconductor and to the possibility to exploit the spontaneous and piezoelectric field of AlGaN/GaN heterostructure to develop HEMTs with a sheet carrier concentration at the interface as high as $2 \times 10^{13}$ cm$^{-2}$.

In order to utilize GaN materials to their ultimate performance a deep understanding of the transport mechanisms are essential. Here the experience gained with GaAs based HEMTs, where phenomena like electron velocity saturation and velocity overshooting are present, may not be of help due to the intrinsic different high field behaviour of GaN. As a matter of fact, experimental results appear to be in contradiction with a situation in which the electron velocity in the channel is saturated: also at high applied voltages, the high

We gratefully acknowledge ESA support under the ATHENA contract 14205/00/NL/PA.

S. Russo is with the Department of Electronic Engineering of Tor Vergata University, Rome, Italy (phone: 00390672597366; e-mail: stefano.russo@roma2.infn.it).

A. Di Carlo, is the Department of Electronic Engineering of Tor Vergata University, Rome, Italy.

field region (E>300 KV/cm) is confined in a small part near the gate contact, while the source to gate electric field is lower then 100 KV/cm. For this reason managing the source to gate region, introducing additional doping or varying its dimension, can prove very useful to increase devices performances.

The importance of S-D distance, in particular on the trans-conductance collapse, has been already pointed out in [3] where is shown how reducing the source to gate resistance $R_{SG}$ it is possible to obtain a more linear behaviour of $gm$. The aim of this paper is to show how the peculiar transport characteristics of the GaN materials, especially regarding its high saturation field, allows to increase the output current values simply managing the contacts distances, without applying additional techniques like field plate or passivation.

For the sake of simplicity, in the following, for simulated device geometries, we will refer the Source to Gate distance, Gate length and Gate to Drain distance by writing the sequence of those values: i.e., 2.0-1.5-2.0 will indicate a device with 2.0 μm Source to Gate distance, 1.5 μm Gate length and 2.0 μm Source to Drain distance.

## II. Devices Structure and Simulation Tool

Our devices are modelled using a two-dimensional self-consistent Monte-Carlo model, which accounts for four conduction valleys and three valence bands [4]. The Poisson equation is solved by applying a multigrid technique [5]. In our model the parameters for the valleys are estimated from band structure calculations [6, 7]. Phonon scattering (acoustic, optical, and polar optical modes), and ionized impurity scattering are accounted for. The ionized impurity scattering is taken into account according to Brooks-Herring formula. The alloy scattering is neglected, since inclusion of this process only causes small changes in velocity-field characteristics for III-V alloys [8].

Calculation including only the AlGaN/GaN polarization charges predict $I_{DS}$-$V_{GS}$ characteristics higher than experimental ones. As shown in [9], the difference between the calculated and experimental dc drain current is due to the polarization-induced surface as well as interface charges. Moreover, the localisation of polarization charges have different effects on the electrical characteristics. While the surface charges act like a floating gate the interface charges induces a strong shift of the threshold voltage and improves the carrier confinement at the heterojunction. In our simulations we set the values of polarization charges densities of $\sigma_{Top}=0.6\times10^{13}$ cm$^{-2}$ at the AlGaN/Air interface, and $\sigma_{Bott}=0.25\times10^{13}$ cm$^{-2}$ at the GaN/Substrate interface, as results of comparisons with experimental results.

As for the AlGaAs/GaAs HEMT, the device consists of a 40 nm unintentionally doped (5 x $10^{14}$ cm$^{-3}$) Al$_{30}$Ga$_{70}$As/GaAs layer, a 10 nm step-doped (5 x $10^{18}$ cm$^{-3}$) Al$_{30}$Ga$_{70}$As/GaAs layer, and a 5 nm unintentionally doped Al$_{30}$Ga$_{70}$As/GaAs spacer layer followed by a 200nm unintentionally doped (5 x $10^{14}$ cm$^{-3}$) GaAs buffer.

The schematic cross section of the two devices sets is given in Figure 1.
All the simulations included an Monte-Carlo ensemble of near 100.000 particles, while output quantities are the result of an average over at least 10ps after the steady state has been reached.

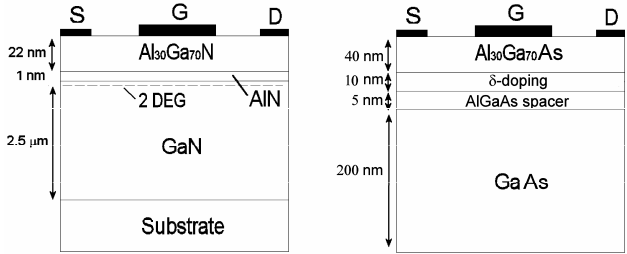

**Figure 1**: schematic representation of the devices cross section.

### III.   RESULTS

As already quoted in the introduction, the electron injection in the channel and, as a consequence, the output current, is ruled by the electric field in the S-G region. Our idea is to increase the electric field in the S-G region varying the S-G distance.

To explain the results shown later, referring to Figure 2, where we represented the electric field along the channel in a 1.0-1.0-1.0 AlGaN/GaN and AlGaAs/GaAs HEMT, we will call $E_{SG}$ the electric field along the channel at the gate edge towards the source contact and $E_{GD}$ the electric field at the gate boundary towards the drain contact.

In order to study the effects of the downscaling of the contact distances we performed two set of simulations for both, GaN based HEMT and GaAs based HEMT. We varied the geometry of the devices keeping the applied voltages fixed: in the first set of simulations we reduced the distance between the Source and Gate contacts, in the second the distance between Gate and Drain. More in details, we observed the devices behaviour for both AlGaN/GaN based and AlGaAs/GaAs based HEMTs, whose geometry has been set x-1.0-1.0 and 1.0-1.0-x, with x ranging between 0.2 and 2.4 μm. The variations of the output current obtained are shown in Figure 3: a downscaling of the S-G distance in GaN HEMT produces an high increase in the output current.

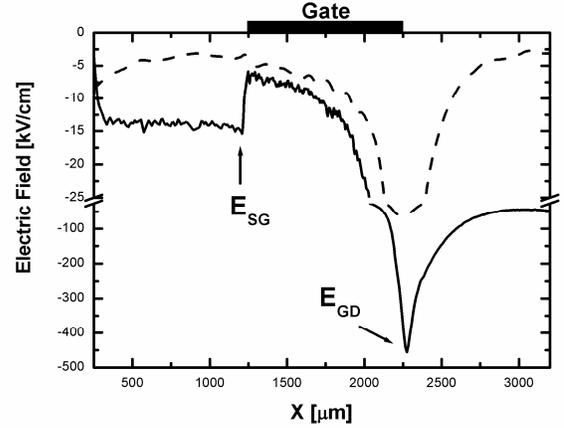

**Figure 2**: Behaviour of the electric field inside an 1.0-1.0-1.0 μm AlGaN/GaN HEMT (solid line) and AlGaAs/GaAs HEMT, with V$_{DS}$= 15V V$_{GS}$= 0V applied voltages for the former and V$_{DS}$= 3V V$_{GS}$= 0V for the latter. The electric field on the boundary of the gate contact has been named E$_{SG}$ and E$_{GD}$.

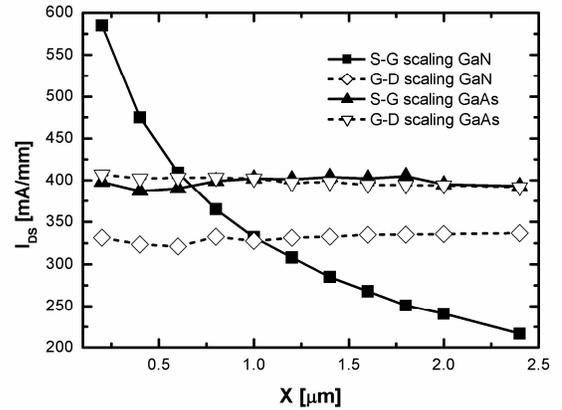

**Figure 3**: Output current of the simulated devices as a function of the length of the downscaling of the S-G region (full lines), and of the G-D region (dashed lines). Square symbols indicate results relatives to AlGaN/GaN HEMTs and triangular symbols indicate AlGaAs/GaAs HEMTs.

Such effect can be explained looking at the differences for the high field transport characteristics of the two materials. In GaAs based devices, scaling the S-G distance results in electric field values of $E_{SG}$ which correspond to saturated electron velocity for all the studied S-G lengths. On the other hand, for GaN based devices, the electric field values span in the ohmic range, that is, where electrons velocity is linearly dependent on the electric field. This implies that small variation of $E_{SG}$ in the GaN based HEMTs results in a high variation of the electrons velocity, while for GaAs based the $E_{SG}$ variation will not produce any visible effects.

The connection existing between the $E_{SG}$ values and the output current in GaN based HEMTs is shown in figure 4.

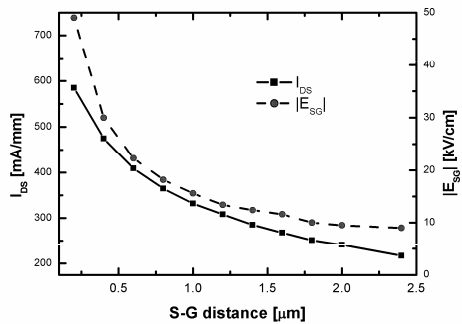

**Figure 4:** Output current of the simulated devices as a function of the length of the downscaling of the S-G region (full line), and the corresponding value of $E_{SG}$ (dashed line).

Reducing the S-G distance, has also good consequence on the trans-conductance values. In fact, the threshold voltage does not depend on the Gate distance, so that, as the output current is increased, the slope of the transcharacteristic is increased, maintaining the same linearity (see Figure 5).

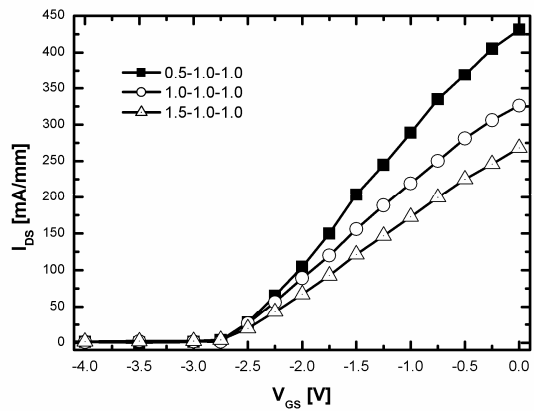

**Figure 5:** Transcharacteristics for AlGaN/GaN HEMTs with different S-G region lengths. The threshold voltages are not changed.

It is also worth noting that the output current increase obtained reducing the S-G distance will not reduce the breakdown threshold as the maximum electric field in the device, that is $E_{GD}$, will stay the same or be even reduced. Moreover, our results indicates that it is possible, keeping the S-G distance fixed, and as a consequence fixing the output current, to design devices with longer G-D distances, whose effect on breakdown voltage has already been shown [10].

## IV. Conclusion

In this paper we presented the results of Monte Carlo simulations, whose aim is to show how with a correct design of the geometry of AlGaN/GaN HEMTS it is possible to enhance devices performances. The results of the simulations indicates an increased field barrier between the Source to Gate region as the Source to Gate distance decrease, allowing an higher output current and higher transconductance. Such results can be obtained without using field plate technique and with technologies accessible now. Moreover, the value of the maximum electric field in the device, which sets the breakdown threshold, is not increased.